\documentclass{aa}
\usepackage{graphicx}
\usepackage{natbib}
\input psfig.tex

\begin{document}

   \title{The Multiplicity Function of Galaxies}
   \subtitle{}

   \author{E. Puddu \inst{1} \and
       E. De Filippis \inst{1,2} \and
           G. Longo \inst{3,1} \and
           S. Andreon \inst{1,5} \and
           R.R. Gal \inst{4}}

   \offprints{E. Puddu; email: puddu@na.astro.it}

   \institute{INAF - Osservatorio Astronomico di Capodimonte, Via Moiariello 16,
    I-80131 Napoli, Italy
    \and
    Astrophysics Research Institute, Liverpool John Moores
    University, Egerton Wharf, Twelve Quays House, Birkenhead CH41 1LD, UK
    \and
    Universit\`a di Napoli Federico II, Via Cinthia, I-80126 Napoli, Italy
    \and
    Department of Physics and Astronomy, Johns Hopkins University,
    Baltimore - MD, USA
    \and
    INAF - Osservatorio Astronomico di Brera, Via Brera 28, I-20121 Milano, Italy 
    }

  \date{Received November 21, 2002; accepted February 5, 2003}
  
  \abstract{
   The multiplicity function (MF) of groups and clusters of galaxies
   is determined using galaxy catalogues extracted 
   from a set of Digitized Palomar Sky Survey (DPOSS) plates. 
   The two different types of structures (of low 
   and high richness) were identified using two different algorithms: a 
   modified version of the van Albada method for groups, and a peak finding 
   algorithm for larger structures. In a $300 deg^{2}$ area up to
   $z<0.2$, we find 2944 groups and 179 clusters. Our MF covers a wide 
   range of richnesses, from 2 to 200, and the two MF's derived by the two 
   algorithms match smoothly without the need for additional conditions or 
   normalisations.
   The resulting multiplicity function, of slope $\alpha= -2.08 \pm 0.07$,
   strongly resembles a power law.
   }

  \maketitle
%

\section{Introduction}
The multiplicity function (hereafter MF), in its differential form,
is defined as the number of groups or clusters per area or volume unit
and per richness unit.\\
The MF, which is the richness spectrum of galaxy aggregates, 
parametrises the observed clustering of galaxies and hence, together with the
correlation and luminosity functions, is one of the fundamental
 cosmological observables.  With respect to the complete
description of clustering, the MF is
complementary to the covariance function (which is related to the
two-point correlation function), being  related to the ratio of the 
amplitude of the higher-order to the two-point correlation functions 
\citep[][ hereafter GT]{got77}.
Due to computational costs and errors, the measurement of correlation
functions of order $N$ becomes unreliable for $N > 3$, and the MF
is therefore a crucial means of obtaining information on higher order 
clustering.\\
The Press--Schechter theory \citep{pressc74} states that the
shape of the mass function (a power-law mass distribution with an exponential
cutoff at the bright end) should provide important clues concerning the
conditions at the
epoch of recombination and does not depend on the cosmic density parameter
$\Omega$. The steepness of the initial density fluctuation spectrum constrains
the broadness of the mass function. \\
The MF, the mass function or the luminosity function all
describe in a similar way the cosmic abundance of objects and, in fact,
present similar shape \citep{bahc79}.

Despite the fact that the early descriptions of galaxy
clustering properties were given in terms of the MF 
\citep{got77}, most authors have focused 
on the shape of the mass function, which can be directly
compared to the PS formalism.
Even when the observed quantity is the MF, some authors \citep{BC93} prefer to
convert it into a mass function using a reliable M/L ratio.
Nevertheless, one must consider all of the uncertainties introduced by
the mass estimation, which are propagated to the mass function determination.
These include errors in the internal velocity dispersion used for dynamical 
mass estimates,
the large intrinsic scatter in the richness-mass relation, and errors in 
assuming dynamical equilibrium for all clusters when using X-ray data 
\citep{gir98}.

The main problem which must be overcome when determining the MF is the 
production of a statistically significant and unbiased catalogue of groups 
and clusters covering a large enough area of the sky and encompassing cosmic 
structures spanning a wide range of richness,
from very low multiplicity structures such as galaxy triplets,
up to very rich clusters with several hundred members.

In the past, catalogues of groups and clusters have been derived from either
3-D data \citep[cf.][]{maia89,ram89,ram01,ram02}, or from projected (2-D) data
\citep{deV75a,deV75b,tur76,mat78,deF99}.
All these catalogues are derived from different data sets and with different
algorithms and are therefore affected by different biases favouring the
detection of structures in a given richness range;
biases induced by the topology of the data, by the limited size of the survey, 
by ambiguities in the selection criteria, etc. \citet{she85} pioneered the 
field of automated cluster finding in optical surveys using peak-finding 
methods, which has been refined and modified in many later projects
\citep{mad90,dal92,lum92,nic01a,gal02}.
Based on a model dependent approach, \citet{Post96} developed
the matched filter technique, which has been widely used, with several variants
\citep{kawa98,SB98,lobo00}, including the adaptive matched filter
\citep{kep99}.
In addition, the availability of multiband high accuracy CCD
data, allowed the implementation of several cluster-finding
methods based on the use of galaxy colours
\citep{GY00,goto02,nic01b,and02}. An independent approach relied on the
Voronoi tessellation technique as a peak finder \citep{ram01,kim00}
and a modified version, taking into account colours was
implemented by \citet{kim02}.
More recently, other, more advanced pattern recognition tools such as
Bayesian clustering \citep{mur02}, maximum likelihood \citep{cocco99}, 
and neural networks (Frattale Mascioli, Priv. Comm.)  have been introduced.

Much less work has  been done to detect poorer structures
such as loose groups; two principal methods (and their successive elaborations)
have been adopted. \citet{tur76} presented the first tentative objective 
identification of groups as enhancements above a reliable threshold in the 
projected galaxy distribution. The "Friends Of Friends" algorithm of
\citet{HG82} generates a measure of correlation among galaxies and their
neighbours, basing on their separation in the full 3-D space.
A noticeable exception to the lack of low-richness catalogs has been  the 
detection of compact groups, where several teams
\citep{dCD95,io99,io02} have proposed different approaches to their
detection.
For the determination of the MF, it is important to note that
its derivation from the above cited catalogues is hindered
by the fact that all of the above algorithms are optimised for the detection 
of either groups {\it or} clusters,
and no systematic work has been done in matching their outcomes in the
transition region between structures of low and high richness.

Here, we attempt the derivation of an accurate MF, starting from
the galaxy catalogues extracted from DPOSS material.

The paper is structured as follows. In Sect. 
\ref{sec:data} we shortly summarise the properties of the 
Digitized Palomar Sky Survey (DPOSS)
data \citep{djo98,djo99,rei91}  used to derive the multiplicity
function described in Sect. \ref{MF}. In Sect. \ref{algo} we
describe the algorithms used to detect groups (Sect. \ref{algogr})
and clusters (Sect. \ref{algocl}), while in Sect. \ref{sec:sim} we
discuss the simulations performed in order to evaluate the
accuracy of the method, expressed in terms of completeness and
fraction of spurious detections, and to evaluate the possible
existence of systematic errors in the ranges of overlapping
richness for the group and cluster finding procedures. Finally, in Sect.
\ref{conclusions}, we draw our conclusions. Through this paper we
assume $H_0 = 100$ $km$ $s^{-1} Mpc^{-1}$.

\section{The data}
\label{sec:data}

The data used in this paper were extracted from the DPOSS photographic
plates \citep{djo98,djo99,rei91}
using the SKICAT package \citep{wei95a} which provides
photometric, morphological and astrometric data for each detected object.
SKICAT also provides a classification (Star/Galaxy) based on a
classification tree \citep{wei95b}.\\
In DPOSS, the three photometric bands ($J, F$ and $N$) are individually
calibrated to the Gunn system \citep{thu76,wad79}
by means of accurate CCD photometry of objects of intermediate luminosity,
(to take into account the nonlinear response of the plates), with preferential
targetting of galaxies.
From the DPOSS data covering the selected regions, we extract, for each individual
object: RA, Dec, total magnitude which best approximates the
asymptotic magnitudes and the object classification.

\begin{figure*}
\centering
\includegraphics[width=13cm]{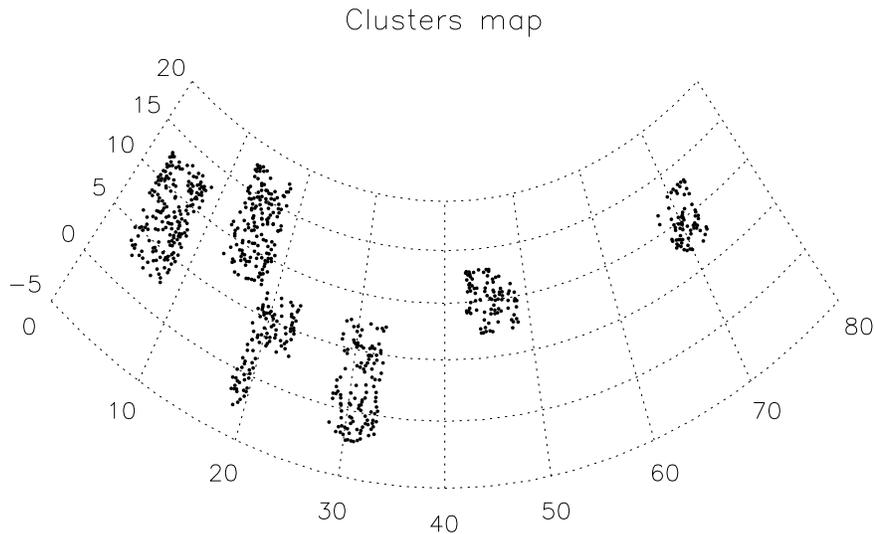}
\caption{Stereographic projection of a transequatorial sky region
(ranging from $\delta=-5$ up to $20$ degs and from $R.A.=0$ up to $80$ degs)
containing the ten selected DPOSS fields listed in Tab. \ref{tabplates}.
In order to trace the global investigated area, all the
detected clusters (without any cut in magnitude) are represented.}
\label{fig:stereo}
\end{figure*}

DPOSS individual plate catalogues must be cleaned
of spurious objects and artifacts (such as multiple detections coming from
extended patchy objects, halos of bright stars,
satellite tracks, etc.). In order to do so, we
mask plate regions occupied by bright, extended and
saturated objects which locally make object detection extremely unreliable.
Subsequently, we matched catalogues obtained in each of the three photometric
bands, by using the plate astrometric
solution and by matching each object in one filter with the nearest
objects in the two other filters 
\citep[with a tolerance box of 7 arcsec, see][]{pao01}.
Due to the different S/N ratios in the three bands, many objects had
discordant star/galaxy classifications in catalogues obtained in the
different bands. The number of such objects obviously increases at faint
magnitudes (it needs to be stressed, however, that this problem is greatly
reduced when a new training set for the classification is adopted, see
\citet{ode02} for details). In order to exclude from our final catalogues
the smallest number of true galaxies,
we discard only the objects classified as stars in all three filters.
Final catalogues were thus obtained for 10 DPOSS plates (see Tab. 
\ref{tabplates}) covering a total area of $\sim 300$ sq. deg.
spread at high galactic latitude ($\|b\|>30 \deg$)
(see Fig. \ref{fig:stereo}), in order to reduce cosmic variance.
Details on the photometric calibration of these particular plates can be
found in \citet{pao01,pao02}. We note that these calibrations are not the same 
as the general DPOSS calibrations described in \citet{gal02}.
Our catalogue of galaxies is limited in magnitude down to the 
 Gunn $r = 20.5 mag$.


   \begin{table}[]
   \caption[]{List of DPOSS plates, from which we extracted our catalogues.
   Notes: (1) calibration from \citet{pao01}; (2) calibration from
\citet{pao02}.}
   \begin{tabular}{cccc}
 \hline
{  Plate  } & {   RA   } & {   Dec   } & {Effective area}\\
{   Num   } & { (2000) } & { (2000)  } & {  ($deg^{2}$) }\\
\hline
610$^{(1)}$ & 01:00 & +15.0 &  30.0\\
680$^{(1)}$ & 00:20 & +10.0 &  37.8\\
682$^{(1)}$ & 01:00 & +10.0 &  38.7\\
688$^{(1)}$ & 03:00 & +10.0 &  38.5\\
693$^{(1)}$ & 04:40 & +10.0 &  30.2\\
752$^{(1)}$ & 00:20 & +05.0 &  32.0\\
755$^{(1)}$ & 00:20 & +05.0 &  24.3\\
757$^{(1)}$ & 01:20 & +05.0 &  24.6\\
827$^{(2)}$ & 01:20 & +00.0 &  10.1\\
829$^{(1)}$ & 02:00 & +00.0 &  24.6\\
\hline

\end{tabular}
   \label{tabplates}
\end{table}

\section{Detection of galaxy overdensities (groups and clusters)}
\label{algo}
Making an arbitrary choice, we use the term ''groups'' to denote those
galaxy aggregates which consist of
less than 20 objects, and ''clusters'' for all richer structures.
This definition is comparable to that of Abell (1958), 
but in our case we set an implicit threshold on the  magnitude difference 
between the brightest and the faintest objects in the same structure, given by
the limiting magnitude. 

\subsection{The procedure for groups}
\label{algogr}
In order to detect galaxy associations of low  richness ($N_{obj}<20$), we have
implemented a modified version of van Albada's algorithm originally
developed for binary systems \citep[see][]{oos89,soares95}.

Taking into account only the position and the apparent magnitude for each galaxy
in our catalog, we first search for the nearest neighbour in a given
magnitude range, and then estimate the probability that the two objects are
physically related.

For the fore/background galaxies, the projected distribution is assumed
to be Poissonian and the probability that the angular separation between a
given galaxy and its nearest neighbour falls
in the range $\theta$ and $\theta+d\theta$, is:
\begin{equation}
P_1 (\theta) d\theta = \exp\left[ - \pi \theta^2 \rho \right] 2 \pi
\theta \rho d \theta\label{eq:p1}
\end{equation}
where $\rho$ is the surface density of background galaxies in the immediate 
neighbourhood.
In order to combine the angular separations of different pairs into a single
distribution, the quantity $x$ is defined as the ratio between the observed
value of the distance ($\theta_1$)
to the nearest neighbour and the expected theoretical mean value $<\theta_1>$
given by Eq.~\ref{eq:p1}:
\begin{equation}
\theta_1 \equiv x \left< \theta_1 \right> =\frac{x}{2 \sqrt \rho}
\end{equation}
The resulting frequency distribution of $x$:
\begin{equation}
p_1 (x) dx = \exp \left[ - \frac{\pi x^2}{4}\right] \frac \pi 2 x dx
\end{equation}
is then independent of the background density $\rho$.  \\

\begin{figure}
\centering
\includegraphics[width=7cm]{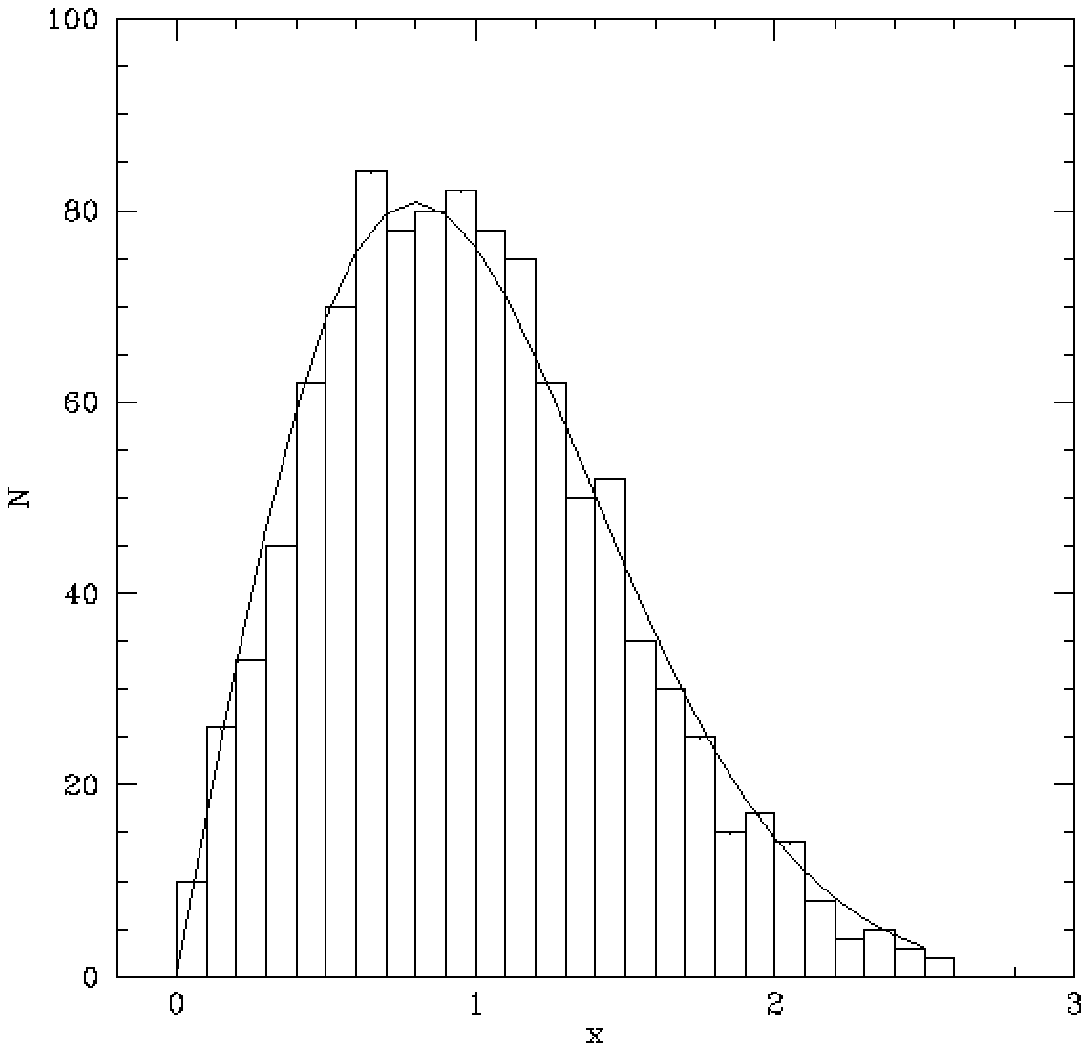}
\includegraphics[width=7cm]{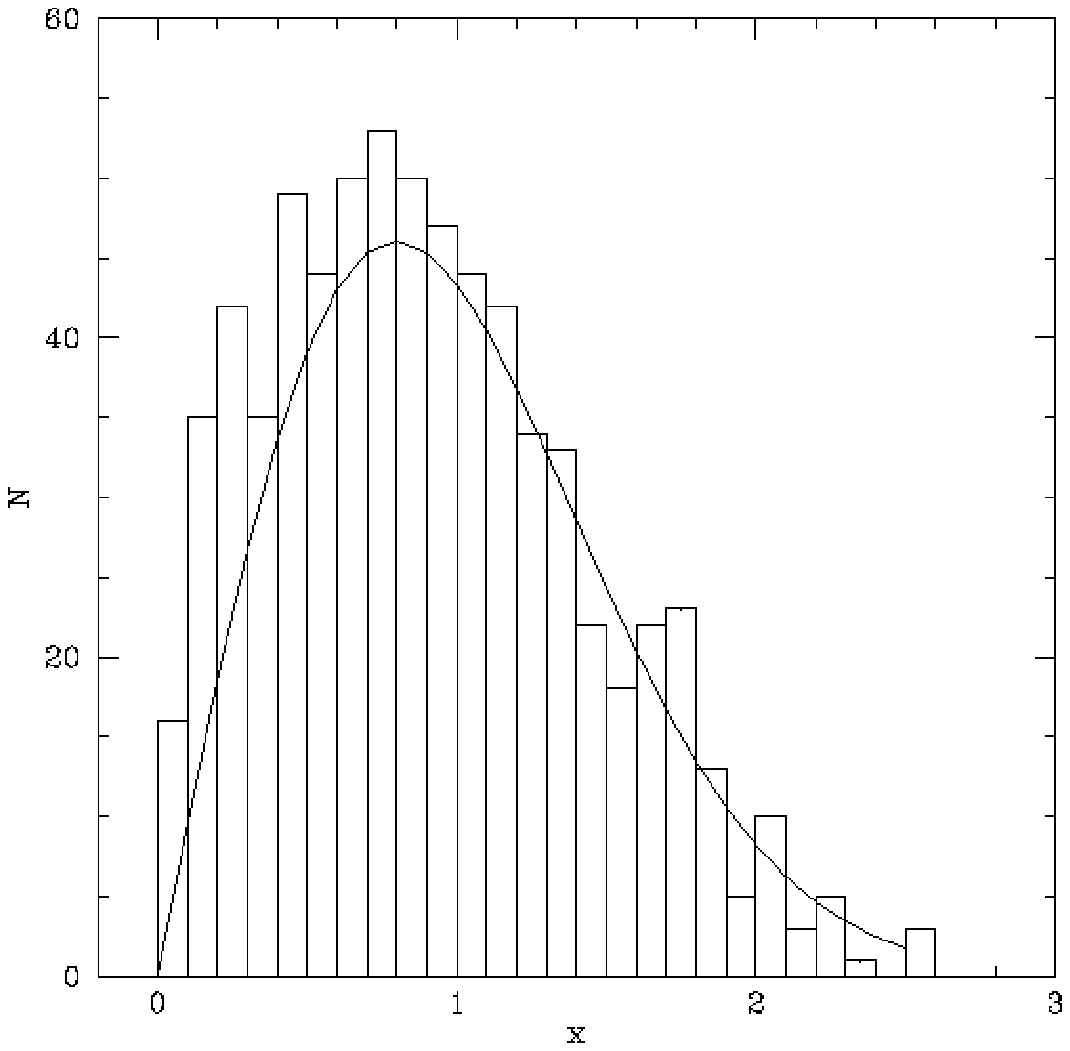}
\caption{Upper panel: comparison of the Poissonian distribution 
(solid line) and the distribution of x's (histogram) in a simulation with 
galaxies randomly distributed in the sky. Lower panel, as upper panel, but
for actual observations: some of the nearest neighbours are physically
linked (related) to the groups and produce an excess of neibourghs at
small x.}
\label{fig:poisson}
\end{figure}

The shape of the observed distribution, $p_0(x)$, and the Poisson
distribution $p_1(x)$, for large $x$, are expected to be similar.
If an excess is found in the
observed distribution  relative to the  Poissonian expectation for small
$x$ (see Fig.~\ref{fig:poisson}, lower panel), it is likely due to physical 
companions, which will tend to cluster at  smaller distances than random 
projections.\\

Normalising the observed distribution to the Poisson
distribution, we can use the excess
$p_0(x)-p_1(x)$, observed at small $x$, to define the probability $p$ that two
galaxies, located at a certain $x$, are physically associated:
\begin{equation}
p \equiv 1- \frac{p_1(x)}{p_0(x)}
\end{equation}

In this formalism, all galaxy pairs having $p$ higher than a given threshold
value are considered to be physical companions.\\

Iteration of the above procedure allows us to estimate the
probability that other companions
of higher order (up to N $\simeq$ 20) are physically related to the first
object by comparing the observed distributions of higher order to the expected
Poissonian distributions (normalised to the local density) for the second,
third, etc. nearest neighbours ($p_2(x)$, $p_3(x)$, etc.).

Groups are then identified by associating all galaxies having probability
$p$ higher than a given threshold value. Groups sharing one or more companions
are finally merged into one single system. The total number of objects defines
our richness for the groups.

To compute the quantity $x$ for every pair of galaxies, it is necessary to
have an accurate estimate of the local galaxy density background $\rho$. To
derive $\rho$, for each galaxy and within each
magnitude interval, one first determines the distance to the $i$-th nearest
neighbour $\theta_{i}$. The relation between $\theta_{i}$ and $\rho$ is given
by the probability that the distance to the $i$-th nearest neighbour
lies between $\theta$ and $\theta+d\theta$:
\begin{equation}
P_{i} (\theta) d\theta = \exp\left[ - \pi \theta^2 \rho \right] \frac{\left(
\pi \theta^2 \rho\right)^{i-1}}{\left( i-1 \right)!} 2 \pi \theta \rho d \theta
\end{equation}
The mean expected value of $\theta_{i}$ is:
\begin{equation}
\left < \theta_{i} \right> = \frac{\Gamma (i+0.5)}{[\left( i-1 \right)! \sqrt
(\pi \rho)]}
\label{eq:back}
\end{equation}
The higher the chosen value of $i$ (i.e. for large distances to the $i$-th
neighbours), the lower the probability of being affected by possible physical
companions, which would lead to an overestimate of the local
background. Furthermore the width of the distribution of the ratio
between $\theta_i$ and its mean value ($<\theta_i>$) decreases with
increasing $i$. Thus if $i$ is large enough, it is possible to obtain
an accurate estimate of $\rho$ from Eq.~\ref{eq:back} by replacing the expected
value $<\theta_i>$ with the observed one $\theta_i$.
On the other hand, $i$ must not be
too large, otherwise too much of the small scale clustering would
be lost, and a large area of the plate will be affected by border effects
(distant companions of galaxies located near the border of the plate will not
follow a Poissonian statistics and will be preferentially located towards the
center of the plate).\\
The choice of the value of $i$ is therefore a
compromise that has to be made by taking into account all ofthe above
factors.\\

\subsection{The procedure of cluster identification}
\label{algocl}
\begin{figure}
\centering
\resizebox{\hsize}{!}{\includegraphics{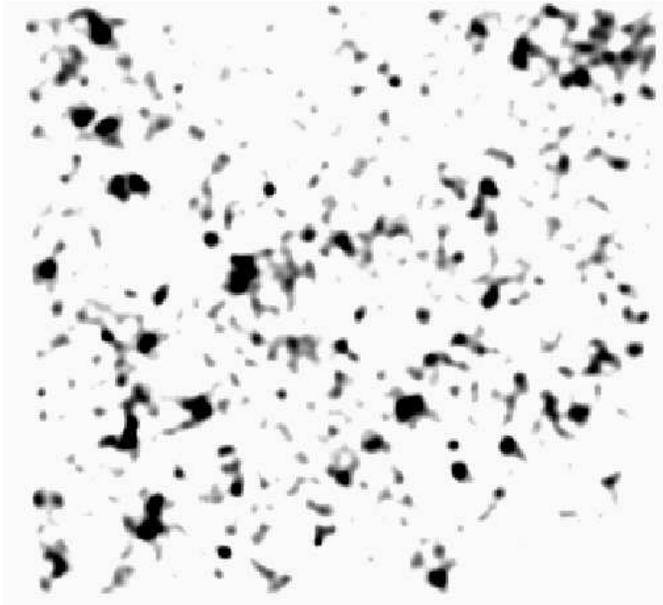}}
\caption{The smoothed two dimensional density map of the number density of
galaxies for a field $5^{\circ}$ x $5^{\circ}$ centered at
$RA = 1h$ and $Dec=+15^{\circ}$ (DPOSS plate n. 610). The smoothing has been
performed by a Gaussian 2-D filter with width ($\sim 3 \arcmin$) typical 
for a cluster core at redshift $0.1-0.2$.}
\label{fig:SEx610}
\end{figure}

Candidate clusters were identified  following a procedure similar to that of
\citet{she85}.
The catalogue of galaxies is binned into equal-area square bins in 
the sky, generating a
two dimensional map (density map) of the number density of galaxies.
The bin size ($1.2 \arcmin$) is chosen such that the mean number of galaxies
per bin is $\sim 1$, in order not to degrade the spatial resolution.
The resulting map (Fig. \ref{fig:SEx610}) exhibits irregular structures 
corresponding to the presence
of overdensities emerging above the intrinsic fluctuations of the background
distribution.

The large fluctuations existing in the distribution of background
galaxies are due to the non-uniform background galaxy
distribution. Once the density map has been created, the analysis of
these maps poses similar problems to those of classical
photometry, so we use S-Extractor \citep{ber96} for the detection
of areas showing enhanced signal. S-Extractor is run on the
density map searching for objects with a minimum detection area of
4 pixels above a global threshold of $0.4$ times the Poissonian
background noise estimated from each plate using a background map.
The evaluation of such background is a crucial step, strongly
affecting the final richness estimate. The use of S-Extractor
poses several problems (which cannot be trivially solved) since it
is optimised to work on images with Gaussian statistics, while in
density maps there are too few objects per bin, and they are
distributed according to Poissonian statistics, thus making
the background determination provided by S-Extractor unreliable.
To circumvent this problem, we were forced to derive the
background map in an alternative way. We first divide the
original density map into sub-images of $\sim 1^{\circ}
 \times 1^{\circ}$, and then compute the Poissonian mean in each box,
subsequently performing a fit with a 2-dimensional polynomial
function of first order. We found a mean background density of 
1640 per sq.deg. with a $\sigma$ of 148 galaxies.
In this way, we remove
those spatial frequencies higher (the clusters) than the mesh
scale length. At the estimated typical redshift in our sample ($z =
0.1 - 0.2$) this scale corresponds to a linear dimension of 9-15
Mpc. This map was then subtracted from the global frame before
running the detection procedure.

The resulting density map was then smoothed in the detection step
using S-Extractor with a Gaussian 2-D filter in order to match the
cluster density profiles and, since we are searching structures
with almost a Gaussian core, the filter width was chosen depending
on the expected average apparent size for the cores ($\sim$ 250
kpc) of clusters in the redshift range ($z = 0.1-0.2$) probed by
our data.
We stress that the choice of the otimal parameters strongly  depends on the
characteristics of the specific data sets
and needs to be tuned on the simulations reproducing the behaviour of true
catalogues.

The extracted parameters characterizing the detected overdensities
are the density centroid in absolute equatorial coordinates (J2000),
the isophotal area above the threshold, the S/N ratio of
detection, and the number of objects inside the isophotal area, which we
use to derive (after the background correction) our richness parameter for 
the clusters.\\


\section{Outlines of the simulation}
\label{sec:sim}
In order to test the limits of our group and cluster detection
procedures, we performed simulations  over a
region having the same area and the galaxy counts as one POSS-II plate.
In this way we could estimate the shortcomings of our
procedure, such as the percentage of spurious detections and the percentage
of lost objects; at the same time this helped in the fine
tuning of the parameters of the detecting algorithms.\\

\subsection{Simulation of the galaxy background}

First we simulated the galaxy background assuming a uniform galaxy distribution.
The number of simulated background galaxies is the average number of
galaxies present in the DPOSS plates (approx $50000$ after excluding all the
galaxies fainter than the limiting magnitude). To each background
galaxy, a sky position, randomly extracted within the plate limits, and an
apparent magnitude, distributed
according to the observed galaxy counts, were assigned.\\

\subsection{Simulation of galaxy groups}
\label{simgr}
\begin{figure}
\centering
\resizebox{\hsize}{!}{\includegraphics{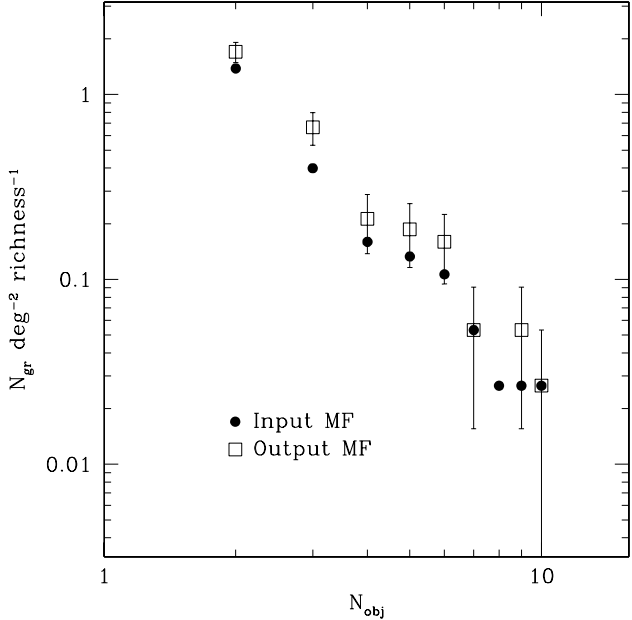}}
\caption{MF of simulated (filled circles) and detected (empty squares) groups.
On the horizontal axis there is the number of galaxies in each group, that is
the richness.}
\label{fig:testgr}
\end{figure}

The number of groups to be simulated was extracted from the multiplicity
function of \citet{tur76}.\\ We began by placing the principal galaxy of each
group at random positions inside each field.
Then, to each principal galaxy we assign an absolute magnitude and a
redshift. Absolute magnitudes were extracted from a Schechter
function with $\mathrm{M^*}=-19.80$ and $\mathrm{\alpha} = -1.25$
\citep{ram99}, while the redshifts were assigned from the galaxy
distribution observed in the Las Campanas redshift survey
\citep{she96}.
To each principal galaxy we then associate a number
of secondary galaxies matching the multiplicity function mentioned
above, each of these galaxies having the redshift
of the corresponding principal galaxy.
Taking into account the estimates provided in the literature,
each simulated group was given a maximum standard dimension depending
on its richness: a maximum radius of $0.26\ \mathrm{Mpc}$ for groups with
$N_{obj} \leq 18$ members, while a maximum radius of $0.55\mathrm{Mpc}$ is used for 
groups with $N_{obj} >18$ members.
All the secondary galaxies belonging to a group were then distributed
inside the group volume, and each assigned an absolute magnitude
generated from the same Schechter function as the brightest galaxies in the
group.
Finally, absolute magnitudes were re-transformed to apparent magnitudes by
taking into account the cluster distance and the average $k$-corrections
from \citet{fuk95}.
The detection algorithm was then applied to the simulated
plates in order to fine tune the algorithm parameters (threshold value of the
probability $p$ and choice of the $i-{\rm th}$ nearest neighbour to compute
the background galaxy density).
The results of the simulations may be summarised as follows:
the group detection algorithm loses 28\% of the simulated groups
 and produces 43\%  spurious detections.\\
Fig.~\ref{fig:testgr} shows that, in spite of the high contamination level,
the MF shape is statistically preserved:
the simulated MF (filled circles) and the detected MF (empty squares) differ
on average by a vertical offset, which we take into account to
correct the final group MF.

\begin{figure*}
\centering
\includegraphics[width=8cm]{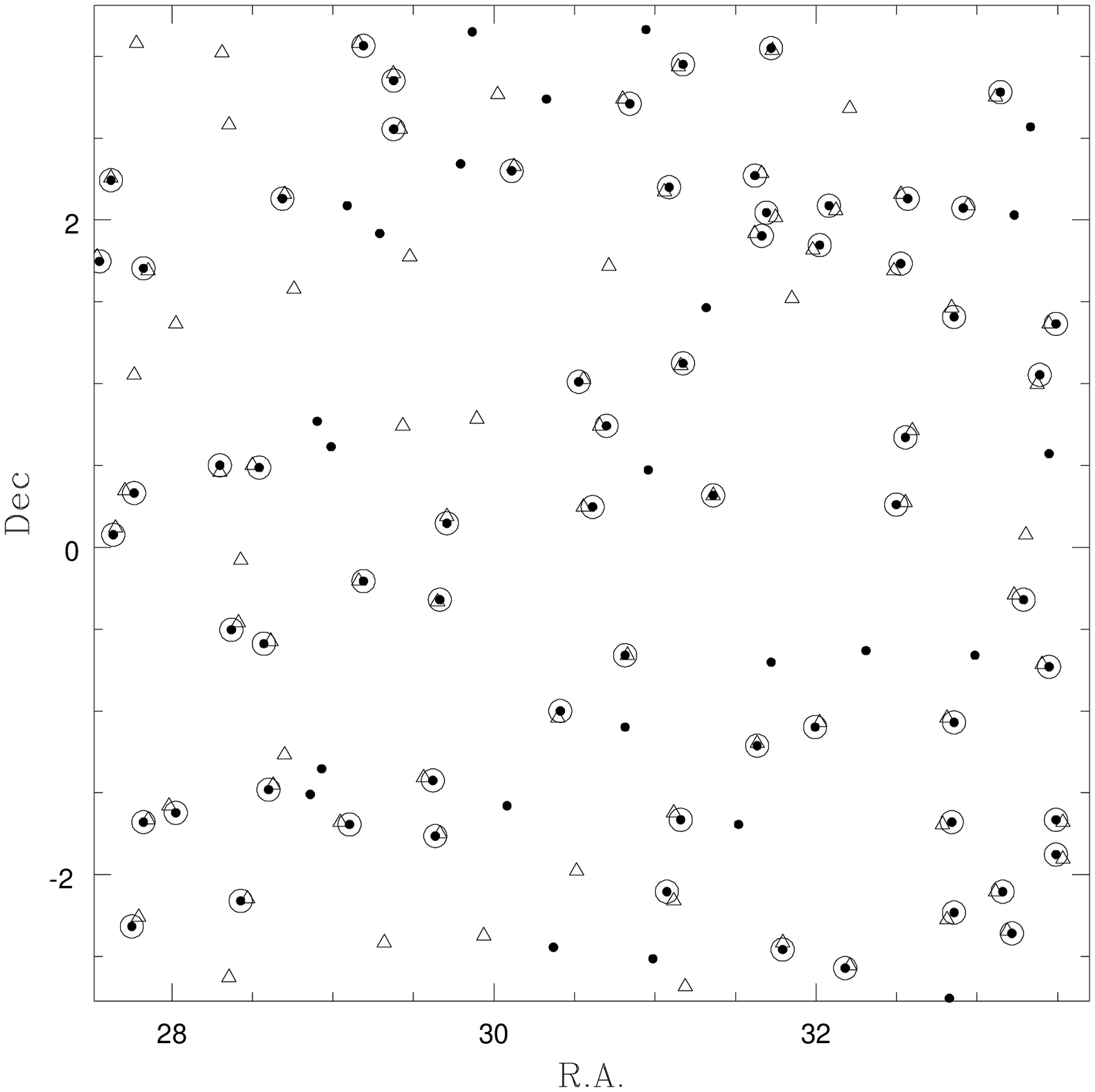}
\includegraphics[width=8cm]{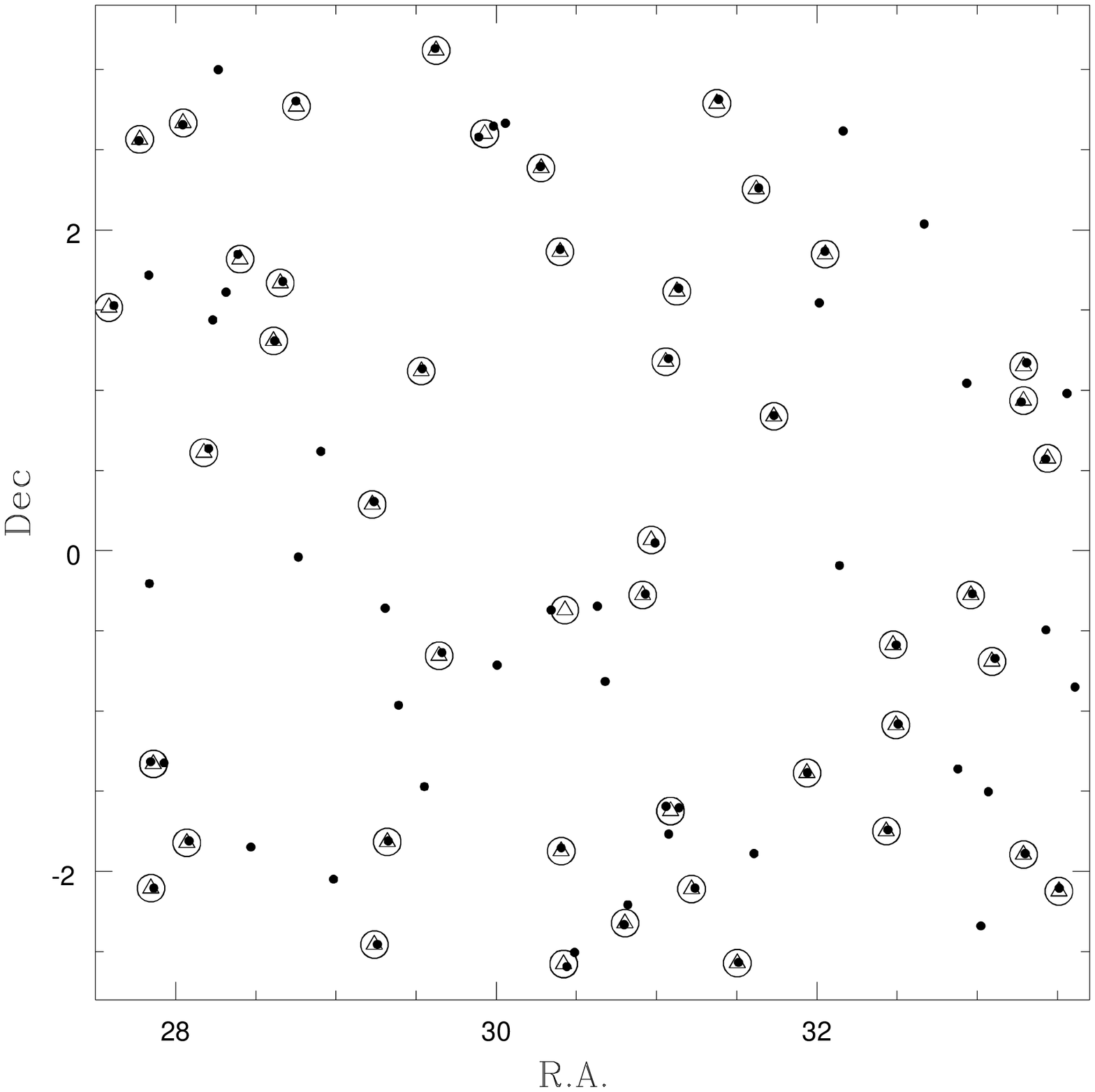}
\caption{Simulated (dots) and detected (triangles) structures.
Left: groups. Right: clusters. Circles highlight
simulated groups/clusters which have been detected.}
\label{fig:simgrcl}
\end{figure*}

In Fig.~\ref{fig:simgrcl} (left panel), we show, as an example, the outcome of 
one typical simulation.
The centers of the simulated (dark dots) and detected (empty triangles) groups
are plotted; a circle is drawn when the two match.\\

\subsection{Simulation of clusters}
\label{simcl}

Cluster simulations were performed with same assumptions used for the groups, 
with some crucial differences.
The number of simulated clusters of a given richness (ranging from 2 to 200
galaxies) in an area of 37.59 squared degrees (approximately the area
of one DPOSS plate) was determined from a preliminary analysis performed
on $10$ DPOSS plates.
In a second step, a power law multiplicity function was used, with
the slope taken from the preliminary multiplicity function.
In this way we tried to take into account the total number of low
richness objects, which could not be measured from our preliminary
analysis.

The absolute magnitudes of the principal galaxies were extracted from a
Gaussian distribution centered on  $-22.99 \pm 1.0\ \mathrm{mag}$
\citep{SGH83}, while those for the secondary galaxies were extracted
from the luminosity function  of \citet{pao01}.
To take into account the richness dependence of the
cluster dimensions,
we arbitrarily adopted a core radius ($\sigma$ of the Gaussian profile)
of $0.5\ \mathrm{Mpc}$ for clusters with $\leq 30$ members, while
a core radius of $1.0\ \mathrm{Mpc}$ was used for clusters with $>30$ members.
Although these values may appear somewhat  high, the adoption of smaller values
for the core radius would only make the detection easier
and therefore the whole procedure more reliable.
As with the groups, the detection algorithm was applied to a large number of
simulated plates to test the algorithm performance as a function of the 
properties of the objects to be detected.

\begin{figure}
\centering
\resizebox{\hsize}{!}{\includegraphics{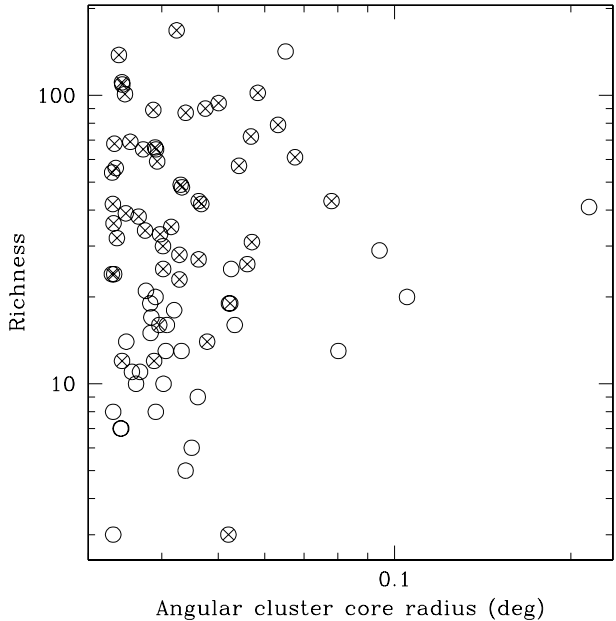}}
\caption{The richness vs. the core radius of the simulated clusters (open
circles). The crosses mark each retrieved cluster.}
\label{Nsig}
\end{figure}

In Fig. \ref{Nsig}, we plot, for a typical simulated plate,
the assigned richness vs. the assigned core radius of the simulated
clusters (open circles) and mark with a cross the clusters
retrieved by the algorithm. Clusters with a very
shallow profile or which are poor are preferentially lost.

\begin{figure}
\centering
\resizebox{\hsize}{!}{\includegraphics{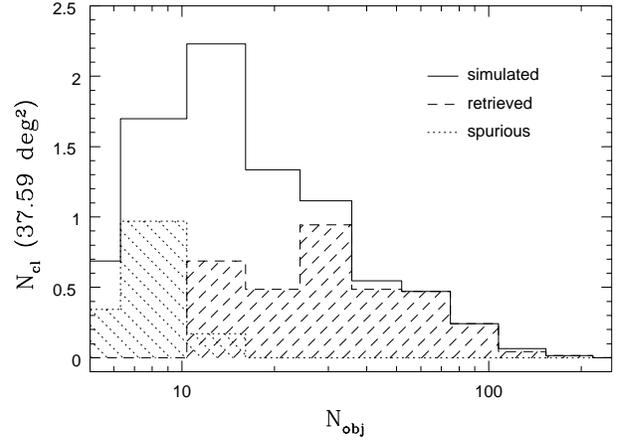}}
\caption{Cluster number in the one plate area is plotted
as a function of estimated richness. The continuous histogram represents
the number of clusters given in input to the simulation;
the dash shaded histogram represents the retrieved clusters
and the dot shaded histogram the spurious detections.
The richness bin grows
exponentially as $2^{n/2}$ (see Sect. \ref{MF}).}
\label{istperc}
\end{figure}

The dependence of the algorithm efficiency on the richness is shown
in Fig. \ref{istperc}, where we plot the number of simulated (continuous
line) and retrieved (dash shaded area) clusters in the typical plate
area.
All but two of the clusters having $N_{obj} > 35$ are retrieved. 
In the range of richness $25<N_{obj} < 35$, 80\% of the clusters are retrieved.
Considering that a cluster belonging to the Abell richness class 0 ($30-49$ 
members in a range of two magnitudes) has $N_{obj}>30$  
($N_{obj}$ includes the cluster galaxies in a range of at least four 
magnitudes), we are complete up to $z=0.2$ at least for all the Abell richness
classes. 
Fig. \ref{istperc} also shows that
spurious detections (dot shaded area) are absent in the richness range where
the algorithm works with the highest efficiency, and occur only in the range 
where the group finder is to be used.

\begin{figure}
\centering
\resizebox{\hsize}{!}{\includegraphics{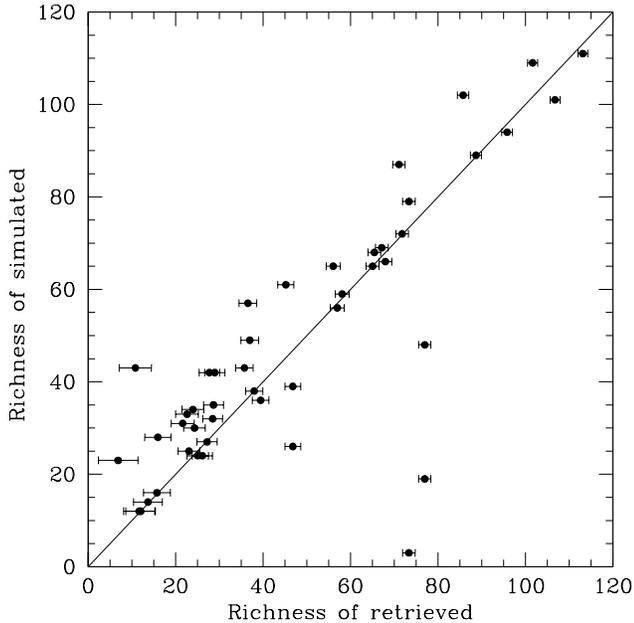}}
\caption{Richness of the simulated vs. the detected clusters.
The errors are inversely proportional to the signal to noise ratio for the
detection.}
\label{rich}
\end{figure}

As already mentioned, the estimate of clusters richness is given by the
number of objects within the detection isophote (isodensity counts).
We wish to stress that this definition of richness depends on the redshift of 
the detected structure.
The quality of the richness estimate has been tested using our simulations.
 In Fig. \ref{rich}, the points follows bisector of the diagram 
 (a bit shifted towards the upper half part of the plot), with a
scatter in richness of $\sim10$ galaxies (which is consistent with the 
background fluctuations).  The small shift indicates an
understimation of the retrieved richnesses. We are comparing the number of 
the galaxies put in a synthetic circular aperture (the simulated) with the 
richness in the isodensity irregular countours, as it is measured in the real 
case: in this way some galaxies are missed. If we use circular apertures 
of the cluster size (which are known in the simulations but not in the actual 
observations), the shift disappears.
Points in the lower right part of the plot are due to
overlapping clusters, for which (in the absence of a deblending procedure) the
richness will obviously be overestimated.\\

\section{The conjoined groups/clusters multiplicity function}
\label{MF}

\begin{figure}
\centering
\resizebox{\hsize}{!}{\includegraphics{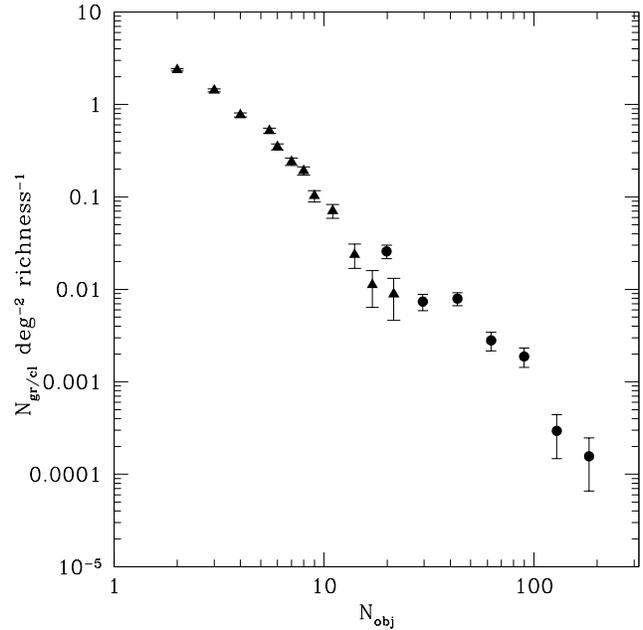}}
\caption{The multiplicity function for galaxy structures ranging from small
groups (filled triangles) to rich clusters (filled circles). We remove 
clusters in the richness range where detection efficiency is low.}
\label{fig:FM}
\end{figure}

Fig. \ref{fig:FM} summarizes our main results.
We plot  the  MF, defined as 
the number of groups or clusters per
unit area and per unit of estimated richness
(the groups/clusters richness is defined respectively in Sect.
\ref{algogr} and Sect. \ref{algocl}). 
For the clusters, the bin grows
exponentially as $2^{n/2}$, in order  to keep the
$S/N$ ratio almost constant along the richness axis.
For the groups, the bin was instead set equal to 1.
In order to exclude the structures detected in the redshift
range where our magnitude limited catalogue is incomplete,
only clusters and groups where the brightest galaxy has $m<16.5$ (in
Gunn $r$) were selected. Assuming that brightest galaxies may be used as
standard candles, our selection in magnitude implies that $z<0.2$.

The procedures described above were applied separately for groups and clusters,
obtaining two different multiplicity functions (marked with different symbols in
Fig. \ref{fig:FM}).  These MFs appear to define a common relation,
without the need for any offsets or normalisations. 
We emphasize that a minor correction for completeness was applied only
to the last point of the clusters MF.
To correct the group's MF for contamination by spurious detections (see
Sect. \ref{simgr}), a global shift derived from the simulations was also
applied.
Only the Poissonian statistical fluctuations have been taken into account in the
error estimate. For the high richness clusters, the error on the richness
estimate is negligible with respect to the bin width. The error becomes relevant
only in the same richness range where incompleteness is also significant.

\begin{figure}
\centering
\resizebox{\hsize}{!}{\includegraphics{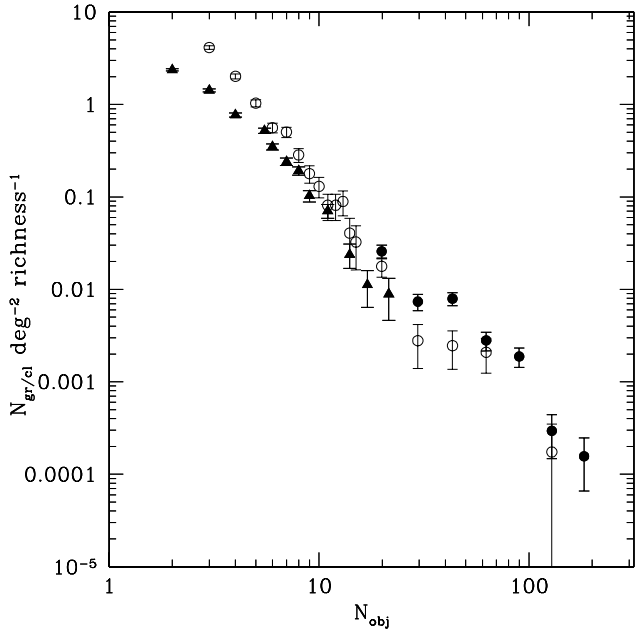}}
\caption{Overplot of the MF of
USGC2 groups (empty circles) on the
multiplicity functions obtained from the DPOSS data. }
\label{fig:FMram}
\end{figure}

In Fig. \ref{fig:FMram} we compare our results with a MF extracted by us from
the USGC catalog of groups \citep{ram02}. We adopt the same representation
scheme for the two data set.
Normalisation to the same volume was applied to the USGC groups, assuming a
uniform distribution of objects in redshift both for our sample and the USGC 
sample. It is important to note that the two catalogs were derived in totally
different ways. The USGC is generated using spectroscopic
redshifts by a percolation method implemented by \citet{ram97} for
group detection, which is designed to reduce the risk of false detections 
introduced by chance projections. 

The agreement between these two MFs (see Fig. \ref{fig:FMram}), derived under 
totally different 
assumptions and using independent data sets, is due to similar biases affecting
the estimated richnesses for both samples. For low $N_{obj}$
structures (groups) the similarity is apparent; in both cases, the methods 
count individual objects fulfilling the respective membership criteria but 
with secondary members having  magnitudes falling within similar (i.e.
$\sim$ four magnitudes) ranges with respect to the primary galaxy. For the
clusters, instead, the different depths sampled by the two data
sets, when compared to the different limiting magnitudes of the
samples themselves, indicates that both methods
sample very similar intervals of the cluster's luminosity
function.

\section{Summary and discussion}
\label{conclusions}
We have implemented two algorithms for the detection of galaxy associations, 
one for groups and one for clusters.
The former is a modified version of van Albada's procedure to
detect galaxy pairs, while the latter the \citet{she85}
approach, which uses a peak-finding
procedure on a density map obtained from the galaxy catalogue.

We evaluated the performance of these methods via extensive simulations,
which show that the group algorithm is reliable up to richness
$20$, and the cluster algorithm is reliable at richnesses above $20$ galaxies.
The two algorithms were then applied to a $\simeq300$ square
degree field extracted from DPOSS data (see Sect. \ref{sec:data}).
The resulting MFs show a remarkable internal consistency from the two procedures
 which produce independent MFs for groups and clusters,
matching with no need for normalisation. Additionally, the MF derived using our
 technique on the 3-D based catalogues of \citet{ram02} agrees with the MF
derived from the projected DPOSS data. The final combined MF is well fit by a
power-law of slope $\alpha=-2.08 \pm 0.07$.
The correlation coefficient on the log-log scale is $-0.98$.\\

The data set we used to determine the MF samples a volume [$\sim
300 deg^{2}$, $z<0.2$] which is slightly smaller than that
explored by \citet{bahc02} [$\sim 400 deg^{2}$,
$z=0.1-0.2$]. The total number of detected structures for $N>10$ in the
\citet{bahc02} and in our sample is respectively $\sim 300$ and $\sim 370$.
 In a forthcoming paper we shall analyze the cosmological implications of the
 derived MF.

\begin{acknowledgements}
The authors wish to thank Marisa Girardi and Michail Sazhin for useful comments and
stimulating  discussions.
\end{acknowledgements}


\end{document}